\documentclass{mem}
\input psfig.sty
\usepackage{natbib}\usepackage{txfonts}\usepackage{balance}
\usepackage{graphicx}
\usepackage[a4paper]{hyperref}
\idline{80}{ 000 }
\def \be {\begin{equation}}
\def \ee {\end{equation}}
\def \bea {\begin{eqnarray}}
\def \eea {\end{eqnarray}}
\begin{document}

\title{Transient cosmic acceleration}

   \subtitle{}

\author{
J. \,S. \,Alcaniz}


\institute{
Observat\'orio Nacional, Rio de Janeiro - RJ, 20921-400, Brasil\\ 
 \email{alcaniz@on.br}
}

\authorrunning{J.S. Alcaniz}

\titlerunning{Transient cosmic acceleration}

\abstract{We explore cosmological consequences of two quintessence models in which the current cosmic acceleration is a transient phenomenon. We argue that one of them (in which the EoS parameter switches from freezing to thawing regimes) may reconcile the slight preference of observational data for freezing potentials with the impossibility of defining observables in String/M-theory due to the existence of a cosmological event horizon in asymptotically de Sitter universes.
 
}

\maketitle{}

\section{Introduction}
Dark energy seems to be the first observational piece of evidence for new physics beyond the domain of the Standard Model of particle physics and may constitute a link between cosmological observations and a more fundamental theory of gravity. Among the many possible candidates for this dark component, the energy density associated with the quantum vacuum or the cosmological constant ($\Lambda$) emerges as the simplest and the most natural possibility. From the observational point of view, flat models with a very small cosmological term ($\rho_{\Lambda} \simeq 10^{-47}$ ${\rm{GeV}}^4$) are in good agreement with almost all sets of cosmological observations, which makes them an excellent description of the observed Universe~(For recent reviews, see \citep{review,paddy,rp,cop,alcanizrev,friedcce}). 

From the theoretical viewpoint, however, the situation is rather different and some questions still remain unanswered. First, it is the unsettled situation in the particle physics/cosmology interface [the so-called cosmological constant problem (CCP)~\citep{weinberg}], in which the cosmological upper bound differs from theoretical expectations ($\rho_{\Lambda} \sim 10^{71}$ ${\rm{GeV}}^4$) by more than 100 orders of magnitude. The second is that, although a very small (but non-zero) value for $\Lambda$ could conceivably be explained by some unknown physical symmetry being broken by a small amount, one should be able to explain not only why it is so small but also why it is exactly the right value that is just beginning to dominate the energy density of the Universe now. Since both components (dark matter and dark energy) are usually assumed to be independent and, therefore, scale in different ways, this would require an unbelievable coincidence, the so-called coincidence problem (CP).

A third question also arises if we try to reconcile the $\Lambda$CDM description of the current cosmic acceleration with the only candidate for a consistent quantum theory of gravity we have today, i.e., Superstring (or M) theory. As is well known, in the standard cosmological scenario, after radiation and matter dominance, the Universe asymptotically enters a de Sitter phase with the scale factor $a(t)$ growing exponentially, which results in an eternal cosmic acceleration. 

In such a background, the cosmological event horizon
\begin{equation}
\Delta = \int_{t_0}^{\infty}{\frac{dt}{a(t)}} \quad \mbox{converges},
\end{equation}  
and this is particularly troublesome for the formulation of String/M theory because local observers inside their horizon are not able to isolate particles to be scattered, which implies that a conventional S-matrix cannot be built. Since the only known formulation of String theory is in terms of S-matrices (which require infinitely separated noninteracting in and out states), we are faced with an important and challenging task of finding alternatives to the conventional S-matrix or, equivalently, defining observables in a string theory described by a finite dimensional Hilbert space~[see \cite{fischler,suss,haylo} for more on this subject]\footnote{It is worth noticing that this problem is not strictly related to $\Lambda$, but a consequence of eternal acceleration.}.

A possible way out of this dark energy/String theory conflict is to construct a dark energy scenario that predicts the possibility of a transient cosmic acceleration. In fact, this possibility can be achieved in the context of the so-called thawing~\cite{pngb,jsa} and hybrid~\cite{hybrid} scalar field models in which a new deceleration period will take place in the future\footnote{\emph{Thawing} models describe a scalar field whose the equation-of-state (EoS) parameter increases from $w \sim -1$, as it rolls down toward the minimum of its potential, whereas \emph{freezing} scenarios describe an initially $w > -1$ EoS decreasing to more negative values~\cite{caldwell,bs,sch}. \emph{Hybrid} models are characterized by EoS that switches from freezing to thawing regimes or vice-versa.}. Examples of transient cosmic acceleration can also be found in brane-world cosmologies~\cite{sahni}, as well as in models of coupled quintessence (dark matter and dark energy), as recently discussed in Refs.~\cite{ernandes,nelson}.

In this short contribution, I will focus on two specific scenarios of transient acceleration that are derived from two different \emph{ans\"atze} on the scale factor derivative of the field energy density~\cite{jsa,hybrid}. Both scenarios are natural generalizations of the exponential potential studied by~\cite{RatraPeebles} and admit a wider range of solutions. 

\section{Models of transient acceleration}

We assume a homogeneous, isotropic, spatially flat cosmology described by the FRW line element $ds^2=-dt^2+a^2(t)(dx^2+dy^2+dz^2)$, where we have set the
speed of light $c = 1$. The action for the model is given by 
\begin{equation}
\label{action}
S=\frac{m^2_{pl}}{16\pi}\int d^4 x \sqrt{-g}[R - {1\over2}\partial^{\mu}\phi\partial_{\mu}\phi-V(\phi)+{\cal{L}}_{m}]\; ,
\end{equation}
where $R$ is the Ricci scalar and $m_{pl}\equiv G^{-1/2}$ is the Planck mass. The scalar field is assumed to be homogeneous, such that $\phi=\phi(t)$ and the Lagrangian density ${\cal{L}}_{m}$ includes all matter and radiation fields.

Now, let us consider the following \emph{ans\"atze} on the scale factor derivative of the field energy density~\cite{jsa,hybrid}
\begin{equation}
\label{ansatz}
\frac{1}{\rho_{\phi}}\frac{\partial\rho_{\phi}}{\partial a}=-\frac{\lambda}{a^{1-2\alpha}} \quad \quad \quad \quad \quad \quad [\mbox{A1}]
\end{equation}
and
\be 
\label{ansatz1}
\frac{1}{\rho_{\phi}}\frac{\partial\rho_{\phi}}{\partial a} = -A
\left(\frac{a^{\kappa-1/2}+a^{-\kappa-1/2}}{2}\right)^2, \quad [\mbox{A2}] 
\ee
where $\alpha$ and $\kappa$ are real parameters, $\lambda$ and $A$ are positive numbers, and the other numeric factors were introduced for mathematical convenience. 

\begin{figure*}[t]
\centerline{\psfig{figure=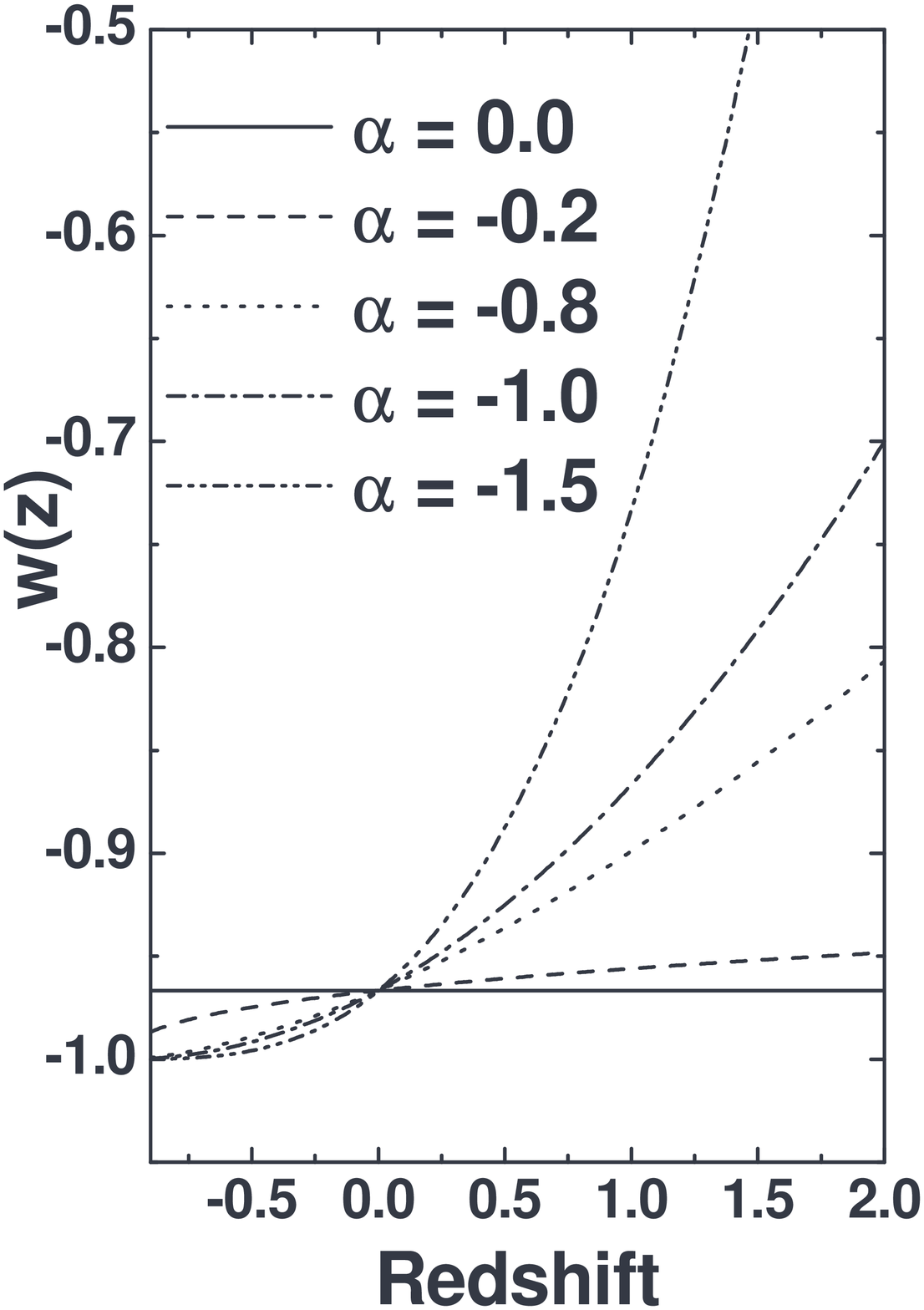,width=2.5truein,height=2.5truein}
\psfig{figure=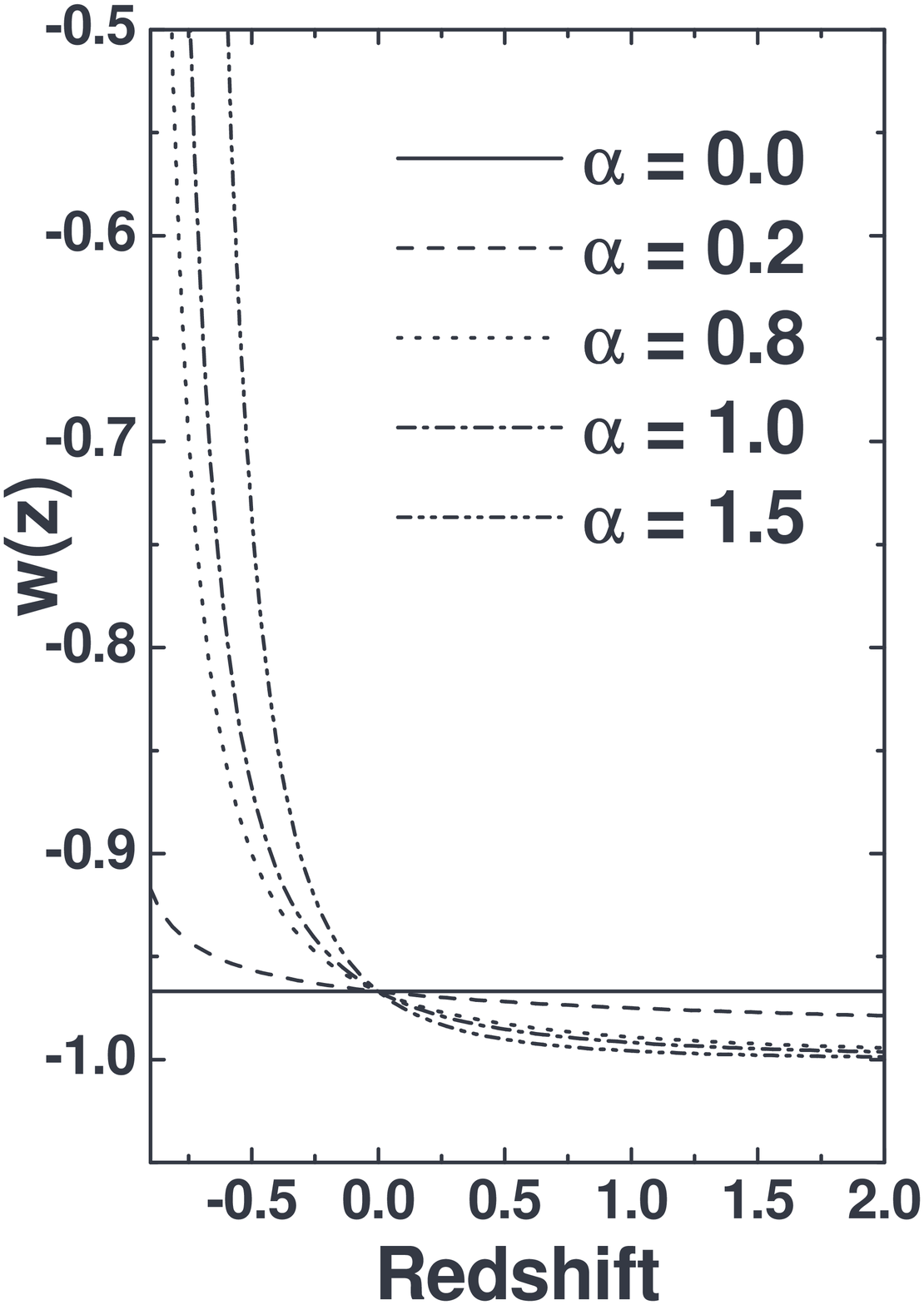,width=2.5truein,height=2.5truein}
\hskip 0.1in} 
\caption{The EoS for the quintessence field derived from A1 (Eq.~\ref{ansatz})  as a function of the
redshift parameter. Note that this \emph{ansatz} provides two different behaviors for $w(a)$, i.e., \emph{freezing} for $\alpha < 0$ (Fig. 1a) and \emph{thawing} for $\alpha > 0$ (Fig. 1b). For this latter case,  $w(a)$ reduces to a constant EoS $w
\simeq -0.96$ [$\lambda = {\cal{O}}(10^{-1})$] in the limit
$\alpha \rightarrow 0$ while for a large interval of $\alpha \neq 0$ it was $-1$
in the past and $\rightarrow +1$ in the future.}
\end{figure*}

By combining the above expressions with the conservation equation for the quintessence component, i.e., $\dot\rho_{\phi}+3H(\rho_{\phi}+p_{\phi})=0$, where $\rho_{\phi}={1\over2}\dot\phi^2+V(\phi)$ and $p_{\phi}={1\over2}\dot\phi^2-V(\phi) $, the expressions for the scalar field and its potential for A1 and A2 can be written, respectively, as
\begin{equation}
 \label{phi1} 
\phi(a) - \phi_0 = \frac{1}{\sqrt{\sigma}}\ln_{1-\alpha}(a)\;, 
\end{equation}
\begin{equation}
\label{gpotential} 
V(\phi)= f_{\alpha}\exp\left[-\lambda\sqrt{\sigma}\left(\phi + {\alpha \sqrt{\sigma} \over 2} \phi^2 \right)  \right]
\end{equation}
and
\bea
\label{phi2}
\phi(a) - \phi_0 &=& {1\over\sqrt{\sigma}}\ln_{\kappa}(a)\;,
\eea
\bea
\label{gpotential}
V(\phi) = f_{\kappa}\exp\left[-\frac{A\sqrt{\sigma}\phi
(1 + \kappa^2\sigma\phi^2)g_{\kappa}(\phi)}
{(\sqrt{1+\kappa^2\sigma\phi^2} + \kappa\sqrt{\sigma}\phi)^2}\right].
\eea
In the above expressions $\phi_0$ is the current value of the field $\phi$, $\sigma$ is a constant,
\begin{equation}
f_{\alpha} = [1-{\lambda\over6}(1+\alpha\sqrt{\sigma}\phi)^2],
\end{equation}
\begin{equation}
f_{\kappa}= 1-\frac{A}{6}\left[\frac{1 - \kappa^2\sigma\phi^2 + \kappa\sqrt{\sigma}\phi g_{\kappa}(\phi)}
{\sqrt{1+\kappa^2\sigma\phi^2} + \kappa\sqrt{\sigma}\phi}\right]^2,
\end{equation}
and $g_{\kappa}(\phi) \equiv \sqrt{1+\kappa^2\sigma\phi^2} + 2 \kappa\sqrt{\sigma}\phi$. The generalized logarithmic functions above are defined as
$$
\ln_{1 - \alpha}(x)\equiv{(x^{\alpha}-1)/\alpha}
$$
and
$$
\ln_{\kappa}(x)\equiv (x^{\kappa} - x^{-\kappa})/2\kappa \quad \quad |\kappa| \leq 1,
$$
and reduce to the ordinary logarithmic function in the limit $\alpha, \kappa \rightarrow 0$. Note that in both limits Eqs. (\ref{phi1})-(\ref{gpotential}) fully reproduce the exponential potential studied in Ref. \cite{RatraPeebles}, while $\forall$ $\alpha, \kappa \neq 0$ the scenario described above represents a generalized model that admits a wider range of solutions (see, e.g., \cite{jsa,hybrid}).

\begin{figure}[h]
\centerline{\psfig{figure=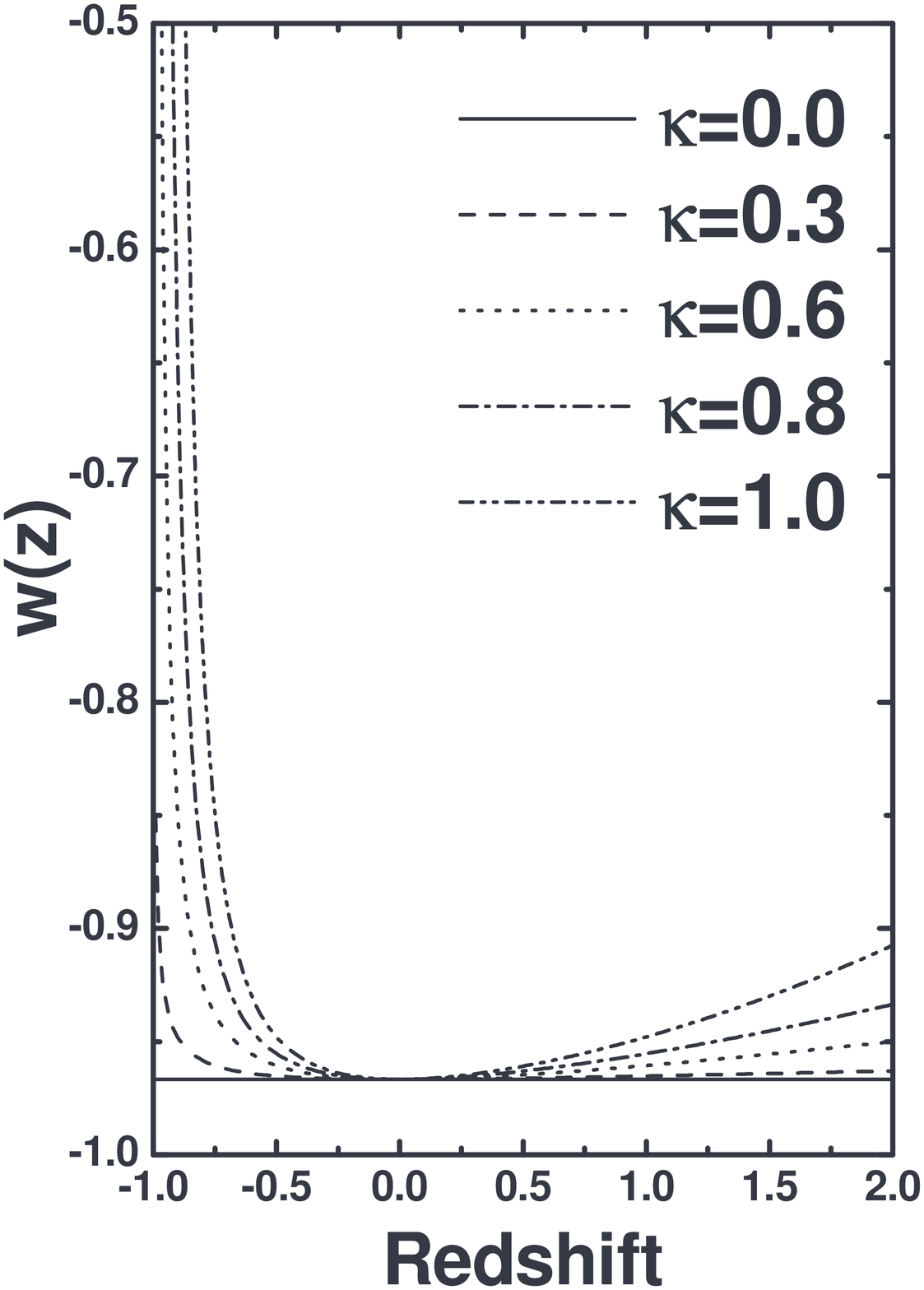,width=2.5truein,height=2.5truein}
\hskip 0.1in} 
\caption{The \emph{hybrid} EoS parameter derived from A2 (Eq.~\ref{ansatz1}) as a function of the redshift. Note that, although behaving as freezing in the past, $w(z)$ becomes thawing as $z$ approaches 0 and will behave as such over the entire future cosmic evolution.}
\end{figure} 
\begin{figure}[h]
\centerline{\psfig{figure=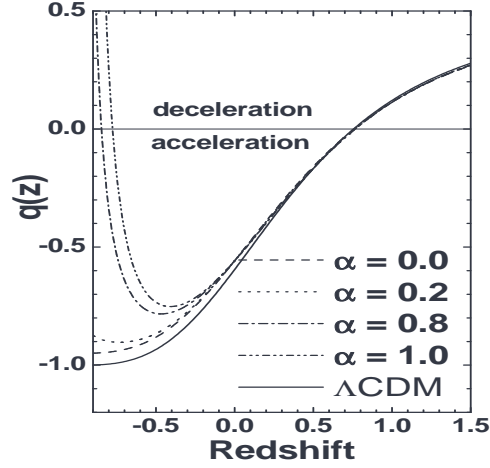,width=2.5truein,height=2.5truein}
\hskip 0.1in} 
\caption{The deceleration parameter for the model A1 (Eq.~\ref{ansatz})  as a function of the redshift for selected values of $\alpha$ and $\Omega_{m,0} = 0.27$. As can be easily seen from this Panel, for some values of $\alpha \neq 0$ the cosmic acceleration is a transient phenomenon. The $\Lambda$CDM case (solid line) is also shown for the sake of comparison.}
\end{figure}

The EoS parameter for these generalized fields can be easily derived by combining the conservation equation for $\phi$ with the \emph{ans\"atze} A1 and A2, i.e.,
\begin{equation}
\label{eq_state} 
w(a)= -1 + {\lambda \over 3}a^{2\alpha}\;, 
\end{equation}
and
\bea
\label{eq_state1}
w(a) = -1 + {A\over 12}(a^{\kappa} + a^{-\kappa})^2\;.
\eea

\begin{figure}[h]
\centerline{\psfig{figure=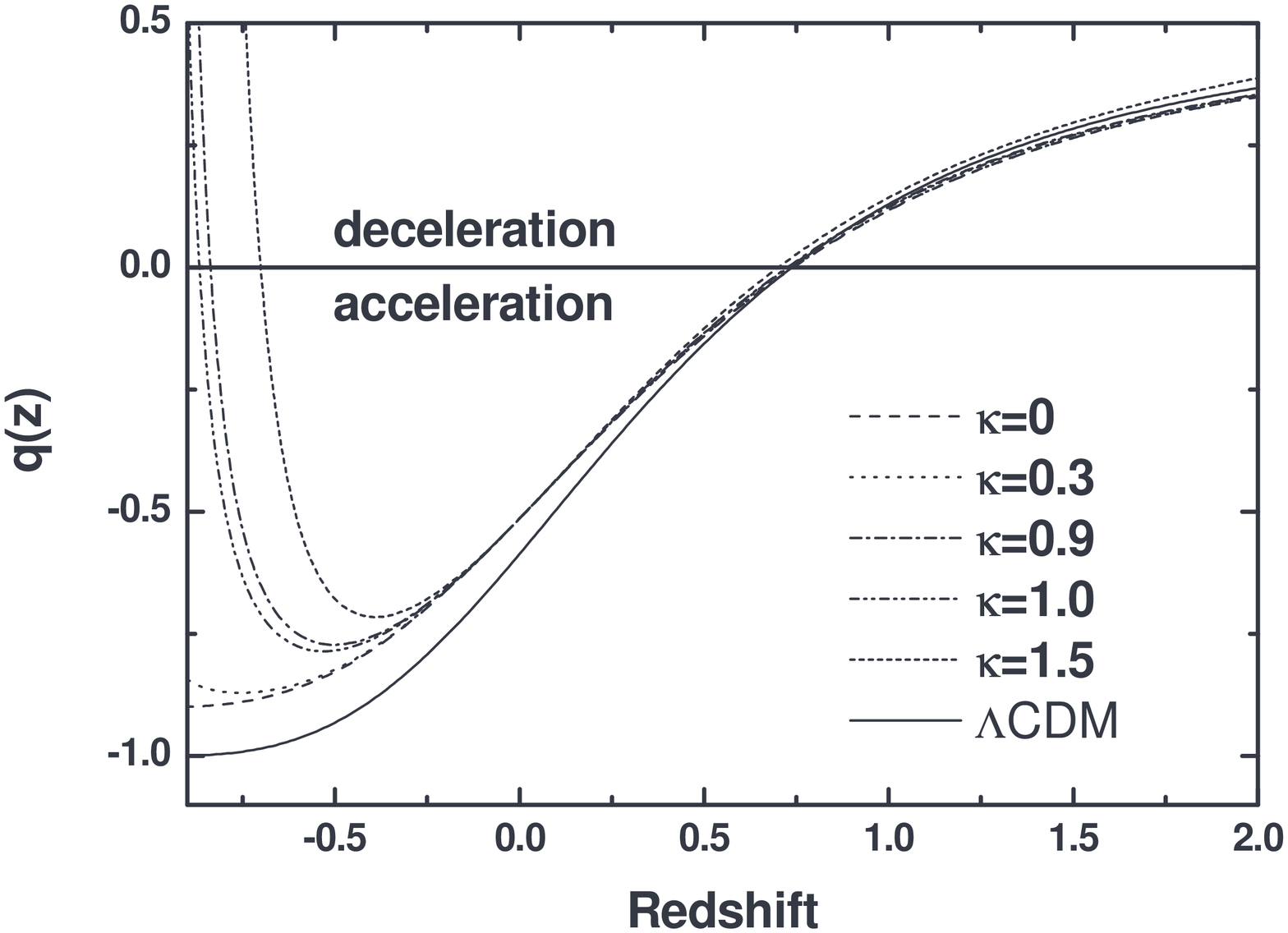,width=2.9truein,height=3.1truein,angle=-90}
\hskip 0.1in} 
\caption{The same as in Fig. 3 for the model derived from A2 (Eq.~\ref{ansatz1}).}
\end{figure}

Figures 1a-1b and 2 show $w$ as a function of the redshift parameter ($z = a^{-1} - 1)$ for some selected values of $\alpha$ and $\kappa$ and $\lambda = A = 10^{-1}$. Clearly, A1 provides two different behaviors for $w(a)$, i.e., \emph{freezing} for $\alpha < 0$ (Fig. 1a) and \emph{thawing} for $\alpha > 0$ (Fig. 1b). In particular, for positive values of $\alpha$, $w(a)$ is an increasingly function of time, being $-1$ in the past, $-0.96$ today, and becoming more positive in the future (0 at $a = 30^{1/2\alpha}$ and 1/3 at $a = 40^{1/2\alpha}$). As mentioned earlier, this is a typical \emph{thawing} behavior, also obtained in the context of Pseudo-Nambu-Goldstone Boson model~\cite{pngb}, whose potential is given by $V(\phi) \propto 1 + \cos{(\phi/f)}$ and EoS parameter approximated by $w(a) = -1 + (1 + w_0)a^F$, where $F$ is inversely related to the symmetry scale $f$.

In the \emph{hybrid} behaviour of Fig. 2, the scalar field EoS behaved as freezing over all the past cosmic evolution, is approaching the value $-1$ today [e.g., it is $w(a_0) \simeq -0.96$ for the above value of $A$], will become thawing in the near future and will behave as such over the entire future evolution of the Universe. Clearly, this mixed behavior arises from a competition between the double scale factor terms in Eq. (\ref{eq_state1}), which in turn is a direct consequence of the generalized logarithmic function used in our \emph{ansatz} (\ref{ansatz1}). Note also that the expressions for A2 are all symmetric relative to the sign of the parameter $\kappa$, which means that one may restrict the $\kappa$ interval to $0 \leq \kappa \leq 1$.

\section{Eternal deceleration}

For a large interval of values for the parameters $\alpha$ and $\kappa$ the behavior of the EoS parameters (\ref{eq_state}) and (\ref{eq_state1}) leads to a transient acceleration phase and, as a consequence, may alleviate the dark energy/String theory conflict discussed earlier. To study this phenomenon, let us consider the deceleration parameter, defined as $q =-a\ddot{a}/\dot{a}^2$ and shown in Figures 3 and 4 as a function of $z$ for some values of $\alpha$ (Fig. 3) and $\kappa$ (Fig. 4) and $\Omega_{m,0} = 0.27$. 

As can be seen from these figures, for some values of $\alpha, \kappa \neq 0$ the Universe was decelerated in the past, began to accelerate at $z_* \lesssim 1$, is currently accelerated but will eventually decelerate in the future. As  expected from Eqs. (\ref{eq_state}) and (\ref{eq_state1}), this latter transition is becoming more and more
delayed as $\alpha, \kappa \rightarrow 0$. In particular, at $a =
20^{1/2\alpha}$ and $a \simeq 9^{1/\kappa}$, $w(a)$ crosses the value $-1/3$, which roughly
means the beginning of the future decelerating phase. A
cosmological behavior like the one described above seems to be in
agreement with the requirements of String/M-theory discussed above, in that the current accelerating phase
is a transient phenomenon. For these scenarios, the cosmological event horizon 
\begin{equation}
\Delta = \int_{t_0}^{\infty}{\frac{dt}{a(t)}} \quad \mbox{diverges},
\end{equation}
thereby allowing the construction of a conventional S-matrix describing particle interactions within the String/M-theory frameworks. A typical example of an eternally accelerating universe, i.e., the
$\Lambda$CDM model, is also shown in Figs. (3) and (4) for the sake of comparison.

\section{Time-dependence of $w(a)$}

Concerning the current observational constraints on the behavior of $w(a)$, it is worth noticing that, although there is so far no concrete observational evidence for a time or redshift-dependence of the dark energy EoS, some recent analyses using current data from SNe Ia, LSS and CMB have explored possible variations in the $w - a$ plane and indicated a slight preference for a freezing behavior over the thawing one~\cite{krauss,trotta}. For instance, \cite{huterer} uses the Monte Carlo reconstruction formalism to scan a wide range of possibilities for $w(a)$ and find that $\sim 74\%$ are for freezing whereas only $\sim 0.05\%$ are for thawing. Similar conclusions are also obtained in Ref.~\cite{trotta} by using the so-called maximum entropy method, where the HST/GOODS SNe Ia data showed $\simeq 1\sigma$ level preference for $w > -1$ at $z \sim 0.5$ with a drift towards $w > -1$ at higher redshifts. 

These results mean that, if such a preference for freezing potentials persists even after a systematically more homogeneous and statistically more powerful data sets become available, the future of the Universe should be an everlasting acceleration toward a de Sitter phase, i.e., in conflict with the String/M theories requirements discussed in Sec. I. This, however, is not the case for one of the scenarios under discussion here (A2) because, differently from pure freezing models, the hybrid EoS given by Eq. (\ref{eq_state1}), although freezing in the past (and, therefore, possibly in agreement with the data), will becoming thawing in the future, so that the phenomenon of a transient acceleration may take place.

\section{Final Remarks}

Motivated by the dark energy/String theory conflict discussed by~\cite{fischler,suss,haylo}, we have explored some cosmological consequences of two scenarios of transient acceleration based on the \emph{ans\"atze} A1 and A2 [Eqs.~(\ref{eq_state}) and  (\ref{eq_state1})]. A basic difference with other quintessence models is that the accelerating phase in these  models does not last forever. After some eons, the equation of state parameter describing the field
component becomes more and more positive with the Universe,
inevitably, returning to an expanding decelerating stage. This predicted transient acceleration, therefore, may be a possible way to reconcile the observed acceleration of the Universe with theoretical constraints from String/M theories.

\begin{acknowledgements}
I am grateful to R. Silva and Z.-H. Zhu for valuable discussions. This work was supported by CNPq - Brazil under Grants No. 304569/2007-0 and No. 481784/2008-0.
\end{acknowledgements}

\bibliographystyle{aa}

\begin{thebibliography}{99}


\bibitem[Sahni and Starobinsky 2000]{review} V.~Sahni and A.~A.~Starobinsky, Int.\ J.\ Mod.\ Phys.\  D {\bf 9}, 373 (2000).

\bibitem[{Padmanabhan 2003}]{paddy}  T. Padmanabhan, Phys. Rept. {\bf{380}}, 235 (2003).

\bibitem[{Peebles \& Ratra 2003}]{rp} P. J. E. Peebles and B. Ratra Rev. Mod. Phys. {\bf{75}}, 559 (2003).

\bibitem[{Copeland et al. 2006}]{cop} E. J. Copeland, M. Sami and S. Tsujikawa, Int. J. Mod. Phys. {\bf{D15}}, 1753 (2006).

\bibitem[{Alcaniz 2006}]{alcanizrev} J. S. Alcaniz, Braz. J. Phys. {\bf{36}}, 1109 (2006).

\bibitem[{Frieman 2008}]{friedcce} J.~A.~Frieman, AIP Conf.\ Proc.\  {\bf 1057}, 87 (2008). arXiv:0904.1832 [astro-ph.CO].


\bibitem[{Weinberg 1989}]{weinberg} S.~Weinberg, Rev.\ Mod.\ Phys.\  {\bf 61}, 1 (1989).

\bibitem[{Fischler {{et al.}} 2001}]{fischler} W. Fischler, A. Kashani-Poor, R. McNees, and S. Paban, JHEP {\bf{3}}, 0107 (2001)


\bibitem[{Hellerman et al. 2001}]{suss} S. Hellerman, N. Kaloper and L. Susskind, JHEP {\bf{3}}, 0106, (2001)


\bibitem[{Haylo 2001}]{haylo} E. Halyo, JHEP {\bf{25}}, 0110 (2001).

\bibitem[{Caldwell and Linder 2005}]{caldwell} R.~R.~Caldwell and E.~V.~Linder, Phys.\ Rev.\ Lett.\  {\bf 95}, 141301 (2005).

\bibitem[{Barger et al. 2006}]{bs} V.~Barger, E.~Guarnaccia and D.~Marfatia, Phys.\ Lett.\ B {\bf 635}, 61 (2006).

\bibitem[{Scherrer 2006}]{sch} R.~J.~Scherrer, Phys.\ Rev.\ D {\bf 73}, 043502 (2006).


\bibitem[{Frieman {{et al.}} 1995}]{pngb} J.~A.~Frieman, C.~T.~Hill, A.~Stebbins and I.~Waga, Phys.\ Rev.\ Lett.\  {\bf 75}, 2077 (1995).


\bibitem[{Carvalho {et al.} 2006}]{jsa}  F.~C.~Carvalho, J.~S.~Alcaniz, J.~A.~S.~Lima and R.~Silva, Phys.\ Rev.\ Lett.\  {\bf 97}, 081301 (2006).

\bibitem[{Alcaniz {{et al.} 2009}}]{hybrid} J.~S.~Alcaniz, F.~C.~Carvalho, Zong-Hong Zhu and R.~Silva, Class. Quantum Grav. {\bf{26}} 105023 (2009). e-Print: arXiv:0807.2633 [astro-ph]

\bibitem[{Sahni and Shtanov 2003}]{sahni}  V. Sahni and Y. Shtanov, JCAP 0311, 014 (2003). 

\bibitem[{Ratra and Peebles 1988}]{RatraPeebles} B. Ratra and P.J.E. Peebles, Phys. Rev D{\bf 37}, 3406 (1988).


\bibitem[{Costa and Alcaniz 2009}]{ernandes} F.~E.~M. Costa and J.~S. Alcaniz, ``Cosmological consequences of a possible $\Lambda$-dark matter interaction,''  arXiv:0908.4251 [astro-ph.CO].

\bibitem[{Fabris et al. 2009}]{nelson} J.~C.~Fabris, B.~Fraga, N.~Pinto-Neto and W.~Zimdahl,``Transient cosmic acceleration from interacting fluids,''. arXiv:0910.3246 [astro-ph.CO]

\bibitem[{Krauss et al. 2007}]{krauss} L.~M.~Krauss, K.~Jones-Smith and D.~Huterer, New J.\ Phys.\  {\bf 9}, 141 (2007).

\bibitem[{Zunckel and Trotta 2007}]{trotta} C.~Zunckel and R.~Trotta,  Mon.\ Not.\ Roy.\ Astron.\ Soc.\  {\bf 380}, 865 (2007).

\bibitem[{Huterer and Peiris 2007}]{huterer} D.~Huterer and H.~V.~Peiris, Phys.\ Rev.\  D {\bf 75}, 083503 (2007).





\end{thebibliography}

\end{document}